\documentclass[conference]{IEEEtran}

\usepackage{booktabs}
\usepackage{paralist}
\usepackage{comment}
\usepackage{graphicx}
\usepackage{multirow}
\usepackage{stfloats}
\usepackage{comment}
\usepackage[font=scriptsize,labelfont=bf]{caption}
\usepackage[labelformat=simple, font=scriptsize,labelfont=bf]{subcaption}
\usepackage{amssymb}
\usepackage{pifont}
\newcommand{\cmark}{\ding{51}}%
\newcommand{\xmark}{\ding{55}}%
\usepackage{paralist}

\begin{document}
%
\title{SIGDROP: Signature-based ROP Detection using Hardware Performance Counters}

\author{\IEEEauthorblockN{Xueyang Wang and Jerry Backer}
\IEEEauthorblockA{Tandon School of Engineering, New York University\\
Brooklyn, NY 11201\\
Email: \{xw338, jerry.backer\}@nyu.edu}
}
\maketitle



\begin{abstract}
Return-Oriented Programming (ROP) is a software exploit for system compromise. By chaining short instruction sequences from existing code pieces, ROP can bypass static code-integrity checking approaches and non-executable page protections. Existing defenses either require access to source code or binary, a customized compiler or hardware modifications, or suffer from high performance and storage overhead. In this work, we propose SIGDROP, a low-cost approach for ROP detection which uses low-level properties inherent to ROP attacks. Specifically, we observe special patterns of certain hardware events when a ROP attack occurs during program execution. Such hardware event-based patterns form signatures to flag ROP attacks at runtime. SIGDROP leverages Hardware Performance Counters, which are already present in commodity processors, to efficiently capture and extract the signatures. Our evaluation demonstrates that SIGDROP can effectively detect ROP attacks with acceptable performance overhead and negligible 
storage overhead.
\end{abstract}

%
\IEEEpeerreviewmaketitle

\section{Introduction}
Return-Oriented-Programming (ROP) is a code-reuse attack approach that allows
an adversary to subvert software control flow and to execute arbitrary (malicious) code \cite{rop}.
In a ROP attack, the adversary constructs the malicious code by chaining sequences of instructions that end with $\mathtt{return}$ instructions (gadgets). ROP attacks can compromise user-level \cite{rop} and kernel-level software modules \cite{rop-kernel}.
\subsection{Motivation}
Customized compiler, address space layout randomization (ASLR) and control flow integrity (CFI) checking are the common solutions used to thwart ROP attacks. 
Customized compiler-based approaches thwart ROP attacks by eliminating gadgets binaries without altering legitimate software behavior \cite{g-free}. Such approaches require access to the software source code which may not be available. ASLR randomly arranges the addresses of the stack, heap, and libraries of a process, preventing the adversary from predicting the addresses of the gadgets and the program stack \cite{aslr}. However, ASLR is vulnerable to information leakage attacks that expose the memory layout \cite{aslr-leakage}. Further it does not protect against just-in-time (JIT) ROP attacks \cite{jit-rop}. In a CFI approach, the software execution flow is compared to a pre-determined golden model that is computed via the software static control flow graph (CFG) or a golden execution model \cite{kbouncer}\cite{ropecker}\cite{ropdefender}\cite{hardware-cfi}.
However, CFI approaches face two main limitations:
\begin{enumerate}
\item \emph{performance and storage overheads}: CFI solutions such as ROPDefender \cite{rop} use binary instrumentation  to monitor software execution and incur performance overheads of up to 3x. Other CFI solutions require up to 19 MB of additional memory to store the golden software model \cite{ropecker}. These overheads are not practical to secure embedded systems with real-time execution requirements or with limited hardware resources.
\item \emph{hardware modifications:} hardware-based CFI solutions such as the one proposed in \cite{hardware-cfi} add new instructions to the processor architecture and are thus not suitable for commodity platforms already in the market.
\end{enumerate}

\subsection{Contribution}
Based on the requirements of a ROP attack, we observe that a ROP payload has the following low-level properties:
\begin{inparaenum}[\itshape a\upshape)]
\item {\bf a sufficiently long chain of gadgets with few instructions in each gadget} \cite{ropecker};
\item {\bf a mispredicted return for each gadget because the target address is not on top of the return address stack}.
\end{inparaenum}
These properties are inherent to any ROP payload and are independent of the monitored program. One can thus use these properties to detect ROP payloads without incurring the limitations of ASLR and CFI approaches.

\begin{table}
\centering
\begin{tabular}{|c|c|c|c|c|}
\hline
 & Low & Low  & \multirow{2}{*}{Portable} & No Source\\  
 & Perf. & Storage & & Code \\ \hline

KBouncer \cite{kbouncer} & \cmark & \xmark & \xmark  & \cmark \\ \hline
ROPecker \cite{ropecker}& \cmark & \xmark  & \xmark & \cmark \\ \hline
ROPDefender \cite{ropdefender}&  \xmark  & \xmark & \cmark & \cmark \\ \hline
G-Free \cite{g-free}&  \cmark &  \cmark & \cmark & \xmark \\ \hline
Hardware-CFI \cite{hardware-cfi}& \cmark &  \cmark & \xmark & \cmark  \\ \hline
SCRAP \cite{scrap}& \cmark &  \cmark & \xmark & \cmark  \\ \hline
{\bf SIGDROP} & \cmark &  \cmark & \cmark & \cmark \\ \hline
\end{tabular}
\caption{{\bf Comparison of SIGDROP to other ROP detection approaches.}}
\label{table:prior}
\end{table}

We propose SIGDROP: a signature-based ROP detection approach that leverages the low-level properties of a ROP payload to detect ROP attacks. SIGDROP uses hardware performance counters (HPCs) readily available in commodity processors to form a runtime signature of the software at different intervals, and notifies a trusted software module when the signature reflects the properties of a ROP payload. We implement a prototype of SIGDROP on the Linux platform. Our implementation and evaluation show that SIGDROP can detect ROP attacks without requiring source code access, customized compiler support and binary rewriting. It is portable to commodity desktops, laptops, and embedded systems without hardware modifications, and has low performance and negligible storage costs. Table \ref{table:prior} compares several recent ROP detection mechanisms to SIGDROP. 

The rest of the paper is organized as follows: in Section \ref{sec:background}, we review ROP attacks and hardware performance counters. We detail the inner workings of SIGDROP in Section \ref{sec:approach} and present its implementation in Section \ref{sec:implementation}. In Section \ref{sec:evaluation}, we evaluate the security effectiveness, performance cost and storage overhead of SIGDROP. We compare our approach to other ROP countermeasures in Section \ref{sec:prior}, and we conclude the paper in Section \ref{sec:conclusion}.

\section{Background}
\label{sec:background}
\subsection{Return Address Stack}
The return address stack (RAS) is a last-in-first-out hardware stack that stores predicted target addresses of $\mathtt{return}$ instructions.
The processor manages the RAS based on the assumption that each $\mathtt{call}$ instruction has an associated  $\mathtt{return}$ instruction (and vice versa). When a $\mathtt{call}$ instruction is fetched, the processor pushes the address of the next instruction on the RAS. When a 
$\mathtt{return}$ instruction is fetched, the processor pops the RAS to predict the target address of the $\mathtt{return}$. The RAS mispredicts the target address due to an overflow or during mis-speculated execution (i.e. miss-peculated branch path has $\mathtt{return}$ instructions).
\subsection{Return Oriented Programming}
The basic idea of ROP is to reuse instructions already residing in memory (e.g. shared libraries, software binary) to induce arbitrary code execution. This allows the adversary to bypass security countermeasures such as data execution prevention that thwart code injection attacks \cite{rop}.

\begin{figure}
	\centerline{\includegraphics[width=0.45\textwidth]{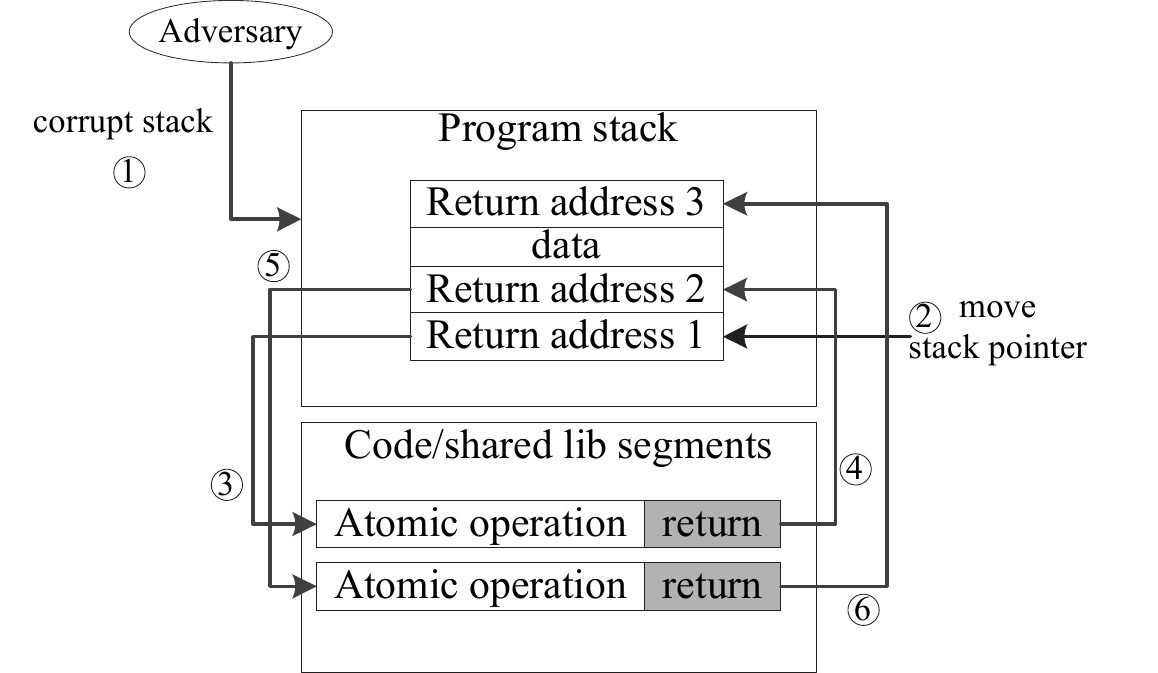}}
	\caption{{\bf ROP Gadget Chaining.} The adversary uses a software vulnerability such as stack-based buffer overflow to control and corrupt the program stack. The adversary then writes the address of each ROP gadget on top of the program stack and moves the stack pointer to the address of the first gadget. Each gadget executed returns to the top of the stack where the address of the next gadget is located. The stack pointer thus servers as the instruction pointer of the ROP payload.}
	\label{fig:rop-high}
\end{figure}

A ROP attack works in two stages: gadget discovery and gadget chaining. During gadget discovery, the adversary searches the memory space for gadgets: sequences of instructions that end with $\mathtt{return}$ instructions. Each gadget performs an atomic operation of the malicious payload (e.g. move value to register or memory, add value to register, make system call).  Figure \ref{fig:rop-high} illustrates the gadget chaining. In this stage, the adversary first finds a software vulnerability such as a stack-based buffer overflow to control and corrupt the software program stack\footnote{The software program stack is different from the hardware-controlled RAS.}. The adversary then writes the addresses of the gadgets on top of the program stack in the order that reflects the execution of the malicious payload.

\subsection{Hardware Performance Counters}
HPCs are a set of special-purpose registers built into the performance monitoring unit of modern microprocessors to store the number of occurrences of hardware activities. HPCs were originally designed for performance debugging of complex software systems. HPCs work along with hardware event selectors which specify the counted hardware events, and the digital logic which increments a counter when a hardware event occurs. Using HPC-based profilers, developers can better understand the runtime behavior of a program and tune its performance \cite{veeperf}.

HPCs provide detailed performance information with much lower overhead than software profilers \cite{wang2013numchecker}. Furthermore, no source code modifications are needed. The hardware events that can be monitored vary from one processor model to another; so does the number of available HPCs. For example, early processors supported very few events and had only a few HPCs. Intel Pentium III has two HPCs and can count about a hundred different events \cite{guide2010intel}. AMD Opteron has over a hundred events and four HPCs \cite{drongowski2008basic}. In contrast, the fourth generation Intel Core Processor has hundreds of events and eight HPCs per core. The ARM Cortex A-15 has six HPCs and can count around 70 events \cite{armrefman}.

\section{Signature-based Detection of ROP Attacks}
\label{sec:approach}
\subsection{Threat Model and Assumptions}
The threat model allows the adversary to leverage a software vulnerability such as a buffer overflow to gain control of the program stack and to launch a ROP attack. The vulnerability can be in kernel or user code, allowing the adversary to launch the ROP attack on privileged or user-level software. We assume the adversary has access to the software binary and shared libraries in order obtain ROP gadgets. The adversary is also able to bypass ASLR via information leakage or via JIT-ROP attacks.

\subsection{High-Level Description of SIGDROP}
\begin{figure}
	\centerline{\includegraphics[width=3.4in]{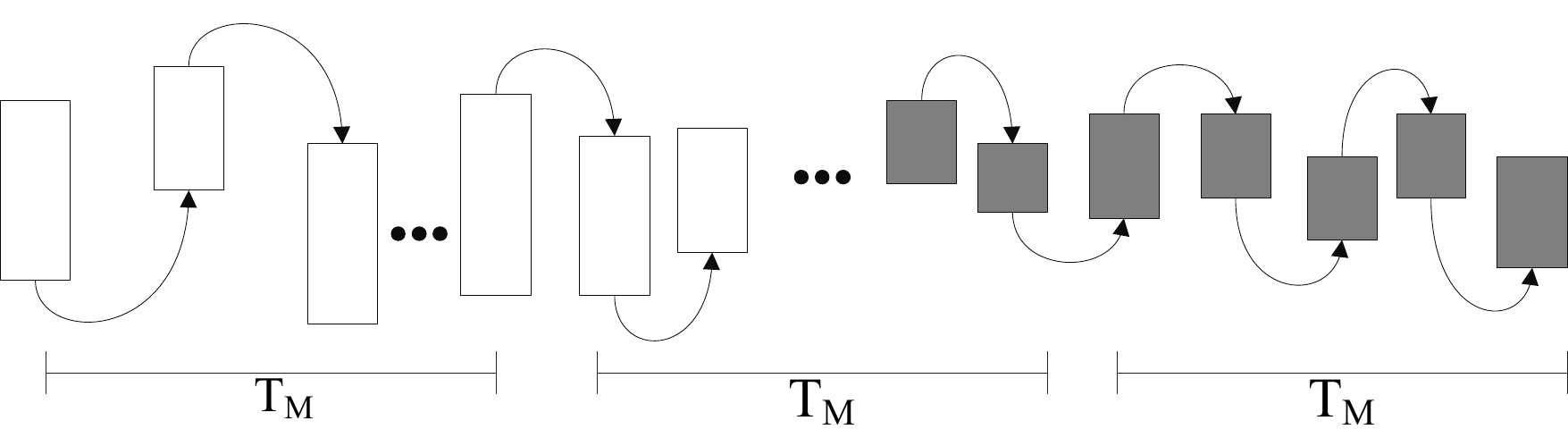}}
	\caption{{\bf High-level description of SIGDROP.} SIGDROP divides the execution a program into monitor intervals of $T_M$ consecutive \textit{mispredicted return instructions}. For each monitor interval, SIGDROP measures the occurrences of \textit{return instructions executed} and \textit{total instructions executed} to determine if there is a sequence of gadgets belong to a ROP payload (shaded rectangles). Each arrow in the figure represents a $\mathtt{return}$ instruction.}
	\label{fig:sigdrop-high}
\end{figure}

ROP payloads have two hardware-level properties:
\begin{enumerate}
\item {\bf mispredicted $\mathtt{return}$ for each gadget}: a ROP gadget ends with a $\mathtt{return}$ instruction, and such a $\mathtt{return}$ will be mispredicted by the RAS because it has no associated $\mathtt{call}$ instruction. Therefore, when a ROP payload is executed, $\mathtt{return}$ instructions are consecutively mispredited.
\item {\bf short gadget size:} to avoid unwanted side-effects on the processor state after its atomic operation, a gadget should have few instructions. Assuming the maximum number of instructions in a ROP gadget allowed is $T_I$, a ROP payload with $n$ gadgets has up to $T_I \times n$ instructions. Recent studies show a gadget in real-world ROP attacks has $T_I \le 6$ \cite{kbouncer, ropecker, scrap}. 
\end{enumerate}
SIGDROP extracts runtime signatures of the monitored program and detects a ROP attack when the signatures reflect the properties of a ROP payload. Figure \ref{fig:sigdrop-high} depicts a high-level description of SIGDROP. When a program is executed, SIGDROP simultaneously monitors three low-level events, \textit{mispredicted return instructions executed}, \textit{return instructions executed} and \textit{total instructions executed}, and takes snapshots of the execution with an interval of $T_M$ consecutive \textit{mispredicted return instructions}. For each snapshot, SIGDROP looks for the following hardware event-based signatures according to the two properties mentioned above respectively: 1) $N_R = T_M$, where $N_R$ is the number of \textit{return instructions} within the monitor interval; 2) $N_I \le (T_I \times T_M)$, where $N_I$ is the number of \textit{total instructions} within the monitor interval.

\begin{figure}
	\centerline{\includegraphics[width=2.5in]{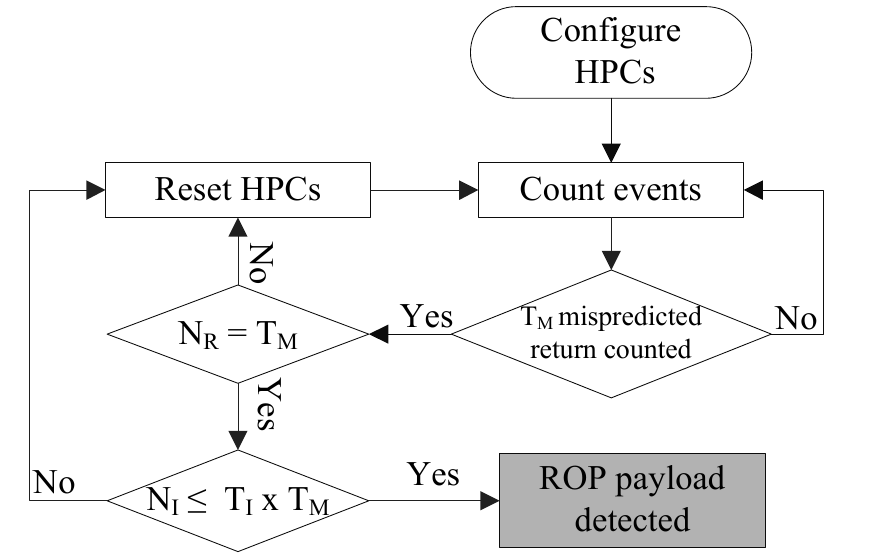}}
\caption{{\bf SIGDROP runtime signature via HPCs}. SIGDROP uses three HPCs to form a runtime signature of the monitored software. SIGDROP counts the 
	 the number of \textit{total instructions}, the number of \textit{return instructions}, and the number of \textit{mispredicted return instructions}. For every $T_M$ \textit{mispredicted return instructions}, SIGDROP checks if the number of \textit{return instructions} $N_R$ is equal to  $T_M$, and if the the number of instructions $N_I$ is less than or equal to $T_I \times T_M$. If both comparisons are true, SIGDROP determines a ROP payload is executed. Otherwise, SIGDROP resets the counters and resumes monitoring.}
	\label{fig:sigdrop-hpc}
\end{figure}

SIGDROP uses three HPCs to measure the occurrences of the monitored hardware events. Figure \ref{fig:sigdrop-hpc} shows the state diagram of SIGDROP with respect to the HPCs. When $T_M$ \textit{mispredicted return instructions} are counted by an HPC, the current execution is suspended and a check is triggered. SIGDROP reads the event counts from the other two HPCs to check if the number of \textit{return instructions} $N_R$ is equal to $T_M$, and if the the number of \textit{total instructions} $N_I$ is less than or equal to $T_I \times T_M$. If both comparisons are true, SIGDROP determines a ROP payload is executed and the execution is terminated. Otherwise, SIGDROP resets all the HPCs and resumes monitoring for the next interval.

\section{SIGDROP Implementation Details}
\label{sec:implementation}
Since the kernel is also vulnerable to ROP attacks, one cannot rely on the HPC values returned by said kernel. Therefore, we implement SIGDROP in a virtualization environment and run the vulnerable kernel and user-level processes in a guest virtual machine (VM). This approach provides the added benefit of detecting ROP attacks at the kernel level. For simplicity of proof-of-concept, we use the same kernel for the host operating system (OS) and the guest VM.

\begin{figure}
	\centerline{\includegraphics[width=2.8in]{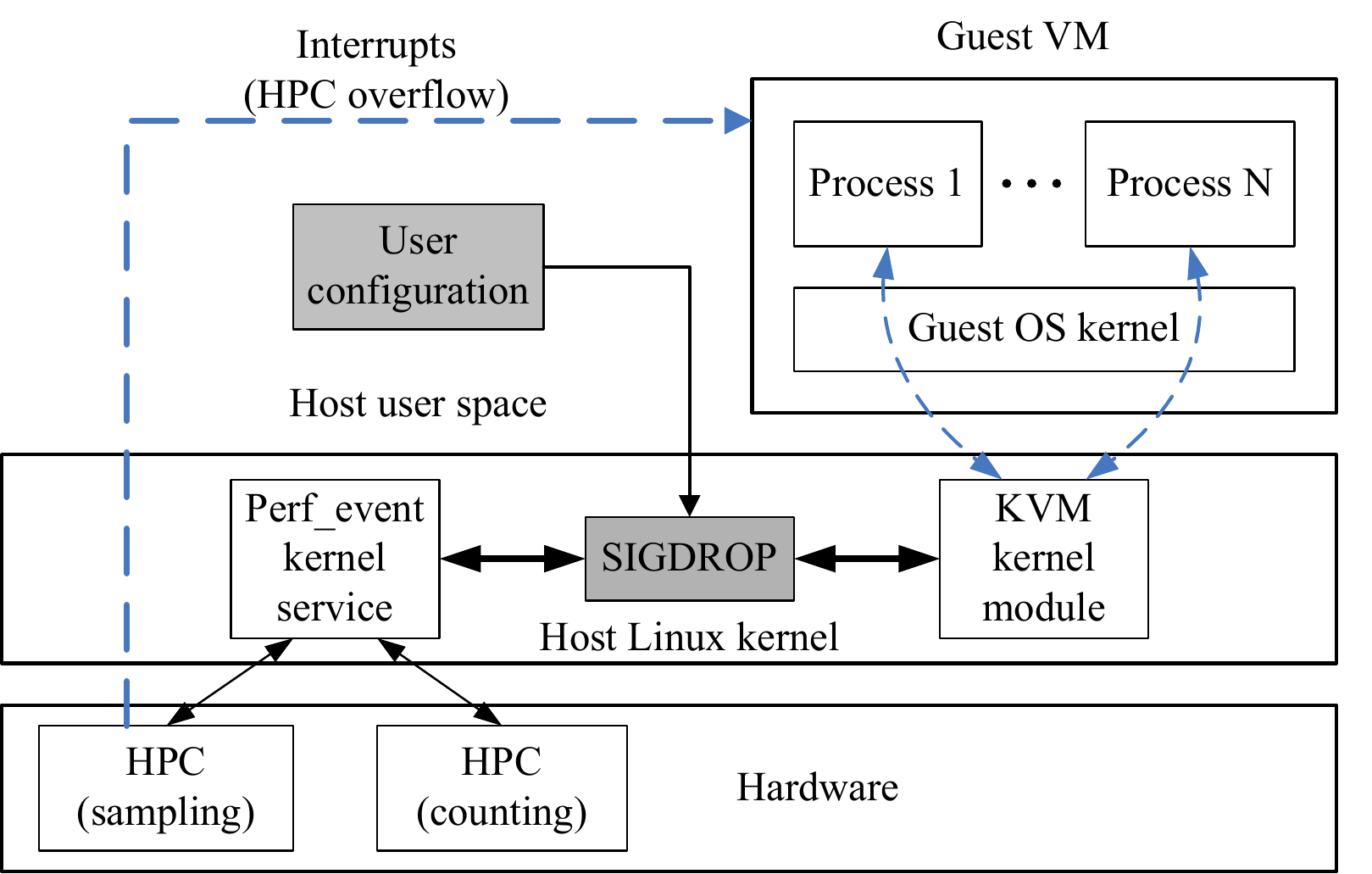}}
	\caption{{\bf SIGDROP high-level architecture.} A SIGDROP module is added into the host kernel to cooperate with the KVM to intercept actions in the monitored guest VM, and communicate with \textit{perf\_event} kernel service to initialize, enable/disable, read, and close HPCs; a program running in the host user space dynamically adjusts $T_M$ and $T_I$ at runtime.
	}
	\label{fig:top}
\end{figure}

Figure \ref{fig:top} illustrates the proof-of-concept implementation of SIGDROP. We use KVM \cite{kvm} to build the SIGDROP virtualization environment. KVM  runs unmodified guest OS and user-level processes (i.e. images) using full virtualization and hardware extensions such as Intel VT for x86 virtualization support  \cite{intelvt}. The core component of SIGDROP is a lightweight module added into the host kernel. The module synchronizes with the host HPC driver and host KVM kernel module to configure the HPCs and to interrupt guest execution when $T_M$ is reached. We use the \textit{perf\_event} HPC driver available in all Linux 2.6+ kernels \cite{perf}, and the KVM kernel module provided in the Linux 3.13 kernel for our proof of concept. \textit{perf\_event} supports two modes of collecting HPC values: counting mode, where the HPC values are aggregated, and sampling mode, where an overflow non-maskable interrupt (NMI) is triggered when the HPC reaches a pre-determined threshold. SIGDROP monitors the number 
of \textit{total instructions} and \textit{return instructions} in counting mode, and monitors the number of \textit{mispredicted return instructions} in sampling mode using $T_M$ as the threshold for an NMI overflow. A program running in the host user space dynamically adjusts $T_M$ and $T_I$ at runtime.

To determine which process has been compromised to launch the ROP attack, SIGDROP cooperates with the KVM host kernel module to monitor every process. When the guest VM is launched, the KVM module sends a unique VM ID (VID) to SIGDROP. SIGDROP then uses \textit{perf\_event} to configure the HPCs to pin the monitoring to the VID, sets the overflow threshold of the HPC for \textit{mispredicted return instructions} to $T_M$, and initializes the HPCs. In this setup, the HPCs monitor all processes that are executed within the guest VM. There are two cases to consider: 
\begin{inparaenum}[\itshape 1\upshape)]
\item no context switches occur during a $T_M$ interval, 
\item context switches occur during a $T_M$ interval.
\end{inparaenum} 
\subsubsection{\bf No Context Switch during $T_M$ interval}
The overflow NMI causes a VM-exit where the processor switches from guest mode to host mode \cite{intelvt}. In addition, KVM updates the VM control structure (VMCS) which is a data structure that stores the state of internal registers of the guest VM \cite{kvm}. If a ROP payload is detected during that interval (as described in Section \ref{sec:approach}), SIGDROP uses the VMCS to obtain the values for the CR3 and EIP register of the guest VM. SIGDROP checks if the EIP is in the range of the kernel virtual memory (i.e. $\mathtt{0xc0000000}$-$\mathtt{0xffffffff}$) which indicates a kernel-level ROP attack. In that case, SIGDROP requests the KVM kernel module to shut down the guest VM. If the EIP range is outside of the guest VM, SIGDROP determines a user-level ROP attack and requests the KVM kernel module to block the process with the CR3 value.
\subsubsection{\bf Context Switches during $T_M$ interval}
The values of the HPCs are accumulated for multiple processes in the guest VM, leading to false positives. Moreover, a smart adversary can induce false negatives by splitting the ROP gadget chain between context switches to other processes. To avoid these issues, SIGDROP uses a lookup table to store the signature of the $T_M$ interval for each user-level process of the guest VM.

A context switch generates a VM-exit that saves the state of the guest VM for the current process, and a VM-entry that loads the state of the guest VM for the next process \cite{intelvt}. The VMCS stores the CR3 of the current process on VM-exit, and the CR3 of the next process on VM-enter \cite{intelvt}. SIGDROP uses these values to update the lookup table entries for the processes. During the VM-exit, SIGDROP uses the CR3 to search the lookup table. If no entry is found, SIGDROP adds a new entry for the CR3 which contains the values of the three HPCs. If an entry is found for that CR3, SIGDROP adds the values read from the HPCs to the current values in the entry. SIGDROP then uses the updated values in the entry to see if they meet the properties of a ROP payload (as described in Section \ref{sec:approach})). During the VM-entry, SIGDROP first resets the values of the HPCs. It then uses the CR3 of the next process to search the lookup table. If no entry is found, SIGDROP sets the 
overflow NMI threshold to $T_M$ and notifies the KVM module to complete VM-entry. If an entry is found, SIGDROP reads the number of \textit{mispredicted return instructions} in the entry, subtracts it from $T_M$, and sets the overflow NMI threshold to the result of the subtraction. This way, when the number of \textit{mispredicted return instructions} for the process accumulates to $T_M$, the overflow NMI is generated and SIGDROP can use the values in the lookup table to detect if the signature of a process matches a ROP payload.

\section{Evaluation}
\label{sec:evaluation}
\subsection{Detection Capability}
To verify the effectiveness of SIGDROP, we perform experiments on a platform with a 3.0GHz Intel Core i5-3330 CPU, which has 11 HPCs on each core. SIGDROP is enable in the host system running 32-bit Ubuntu 14.04, and ROP attacks are launched inside the guest VM running the same OS.

\subsubsection{Setting Parameters}
In SIGDROP, $T_M$ and $T_I$ are critical parameters for ROP detection which need to be properly tuned to distinguish the gadget chains of ROP payloads and those of normal execution flows. A recent research \cite{ropecker} shows that existing real-world ROP attacks have at least 17 gadgets, and the length of longest gadget chain of normal execution flows is 10. The threshold for the gadget chain length $G$ can be a number from 11 to 16 to reduce the false positive and false negative. In our experiment, we choose $G=12$ as the threshold. To avoid false negatives, the monitor interval $T_M =\lfloor\frac{G}{2}\rfloor$, which is 6. For the threshold $T_I$, recent researches on ROP detection \cite{ropecker, scrap} and existing tools for automatic gadget discovery \cite{rop,drop} limit the gadget size to at most 5-6 instructions. In our experiment, $T_I$ is set to 6.

\subsubsection{Crafting ROP Payloads}
In the experiment, we use a small program with a stack buffer overflow vulnerability that can be exploited by applying a long input parameter. The program is compiled with statically linked C libraries. A gadget search tool ROPgadget \cite{ROPgadget} is used to analyze the compiled binary file and generate usable gadgets. We refer to a set of Linux x86 shellcode from Exploits Database \cite{exploit-db} and then chain the found gadgets together to craft ROP payloads that can perform the same actions, such as starting a shell and changing the access permissions. 

\subsubsection{Detecting ROP Attacks}
\begin{figure}
	\centerline{\includegraphics[width=3.4in]{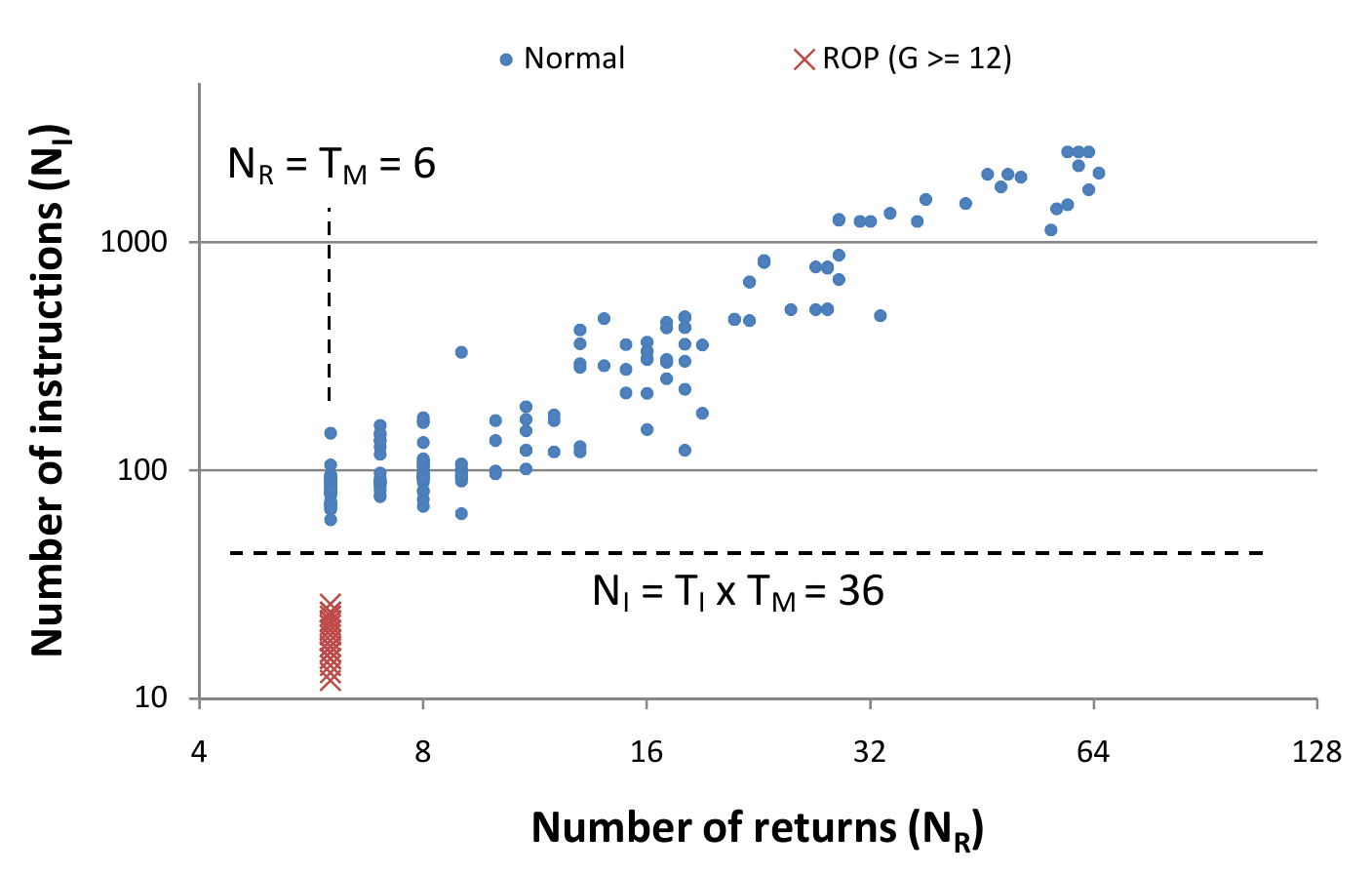}}
	\caption{SIGDROP detection capability. With $T_M = 6$ and $T_I = 6$, all the ROP payloads with the gadget chain length $G \ge 12$ are successfully detected by SIGDROP without any false positive.}
	\label{fig:det-cap-1}
\end{figure}

We test SIGDROP against the ROP payloads as well as normal applications from Linux \textit{/bin/} and \textit{/usr/bin/} and the SPEC INT2006 benchmark suite \cite{spec}. We choose 40 normal applications and each of them is executed 5 times so there are totally 200 executions. We then run 30 ROP payloads with the different gadget chain length $G \ge 12$. Each execution is plotted as a point in Figure \ref{fig:det-cap-1}. The x-coordinate of a point indicates the smallest $N_R$ in a monitor interval during the whole execution while the y-coordinate indicates the $N_I$ in the same monitor interval. From the results we can see that for most of executions of the normal applications, the smallest $N_R$ in a monitor interval is larger than $T_M$, which violates the 1$^{st}$ property of ROP payloads. For those with the smallest $N_R$ in a monitor interval equals to $T_M$, the $N_I$ in the same interval is greater than $T_I \times T_M$, which violates the 2$^{nd}$ property of ROP payloads. Therefore all the 200 executions are correctly considered as normal by SIGDROP. In contrast, for all the executions of ROP payloads, the smallest $N_R$ counted in a monitor interval equals to $T_M$ and the corresponding $N_I$ is less than $T_I \times T_M$, indicating successful detections.

\begin{figure}
	\centerline{\includegraphics[width=3.4in]{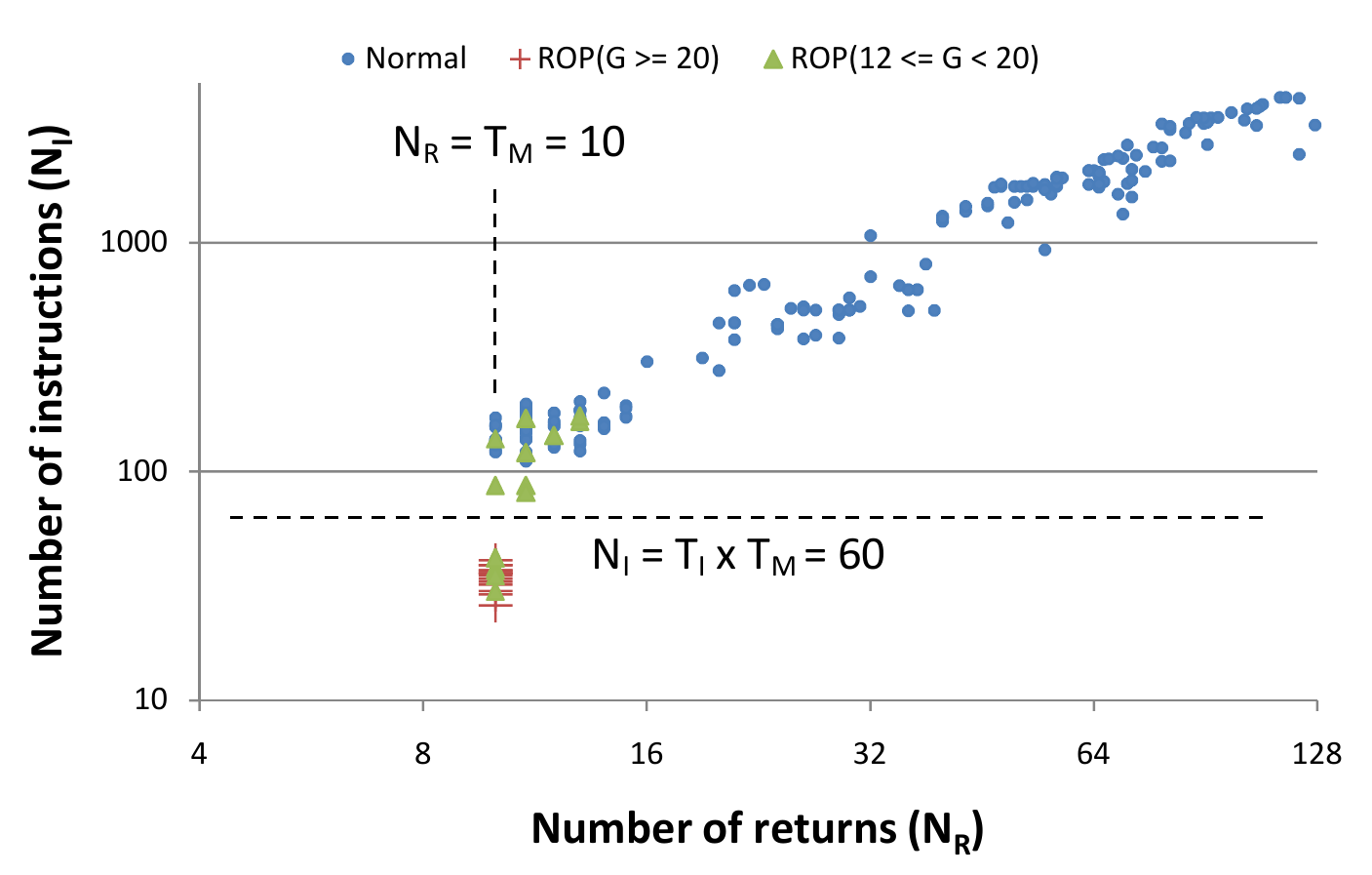}}
	\caption{When $T_M$ is increased to 10, the false positive rate remains zero but there are false negative encountered for detecting ROP payloads with the gadget chain length $12 \le G < 20$.}
	\label{fig:det-cap-2}
\end{figure}

\subsubsection{Detection with a Larger $T_M$}
In the next test, we increase the monitor interval $T_M$ to 10, and test SIGDROP against the same set of normal applications and ROP payloads. The results are illustrated in Figure \ref{fig:det-cap-2}. We can observe that the false positive rate remains zero for the normal applications but there are some false negative encountered when the gadget chain length $12 \le G < 20$. The results confirm that to guarantee zero false negatives, $T_M$ has to be less than $\lfloor\frac{G}{2}\rfloor$. However, a smaller $T_M$ results in higher performance overhead. There is a trade-off between the detection accuracy and the system performance. 

\subsection{Performance Evaluation}
Our next experiment is to test the performance overhead in the guest system when SIGDROP is enabled. The evaluated platform has the same hardware configuration the previous experiment. The host is running 32-bit Ubuntu 14.04 (kernel version 3.13.1) with 8GB RAM and 4-core configuration; The guest VM is running the same OS with 2GB RAM and 1-core configuration.

\begin{figure}
	\centerline{\includegraphics[width=3.4in]{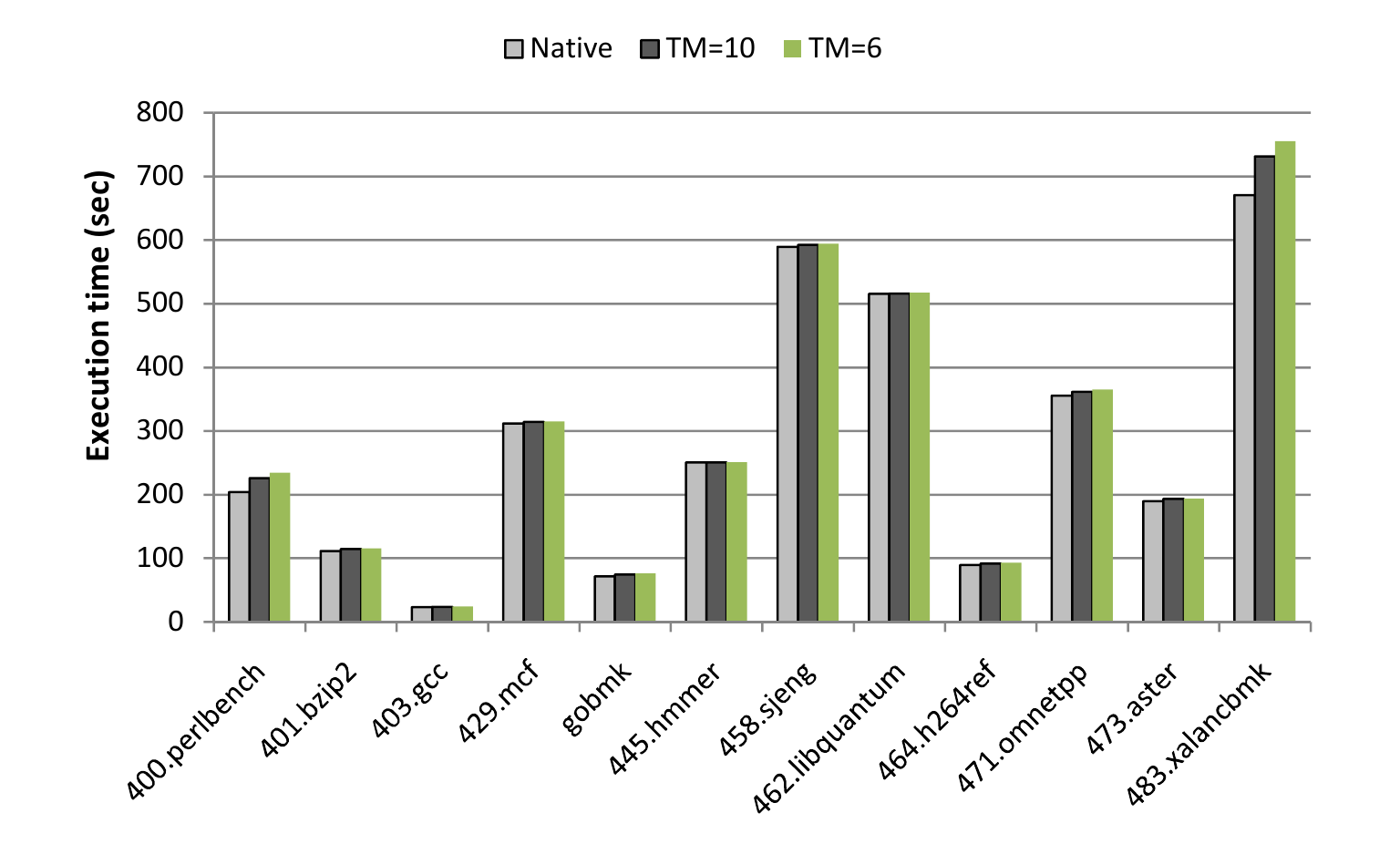}}
	\caption{The SPEC INT2006 benchmark results: CPU computation overhead of a guest VM when SIGDROP is enabled with $T_M=10$ and $T_M=6$.}
	\label{fig:spec}
\end{figure}

We choose two benchmarks to evaluate the overhead in the guest VM on CPU computation and system throughput:

{\bf SPEC CPU Benchmark} The SPEC INT2006 benchmark suite is used to evaluate the computation performance in the guest VM when SIGDROP is enabled with the monitor window size $T_M$ of 10 and 6, respectively. The results are compared to the normal guest VM performance without SIGDROP, as illustrated in Figure \ref{fig:spec}. All the numbers are averaged over 20 runs. The average overhead of the guest computation performance is 3.18\% when $T_M=10$ and 4.67\% when $T_M=6$.

\begin{figure}
	\centerline{\includegraphics[width=3.4in]{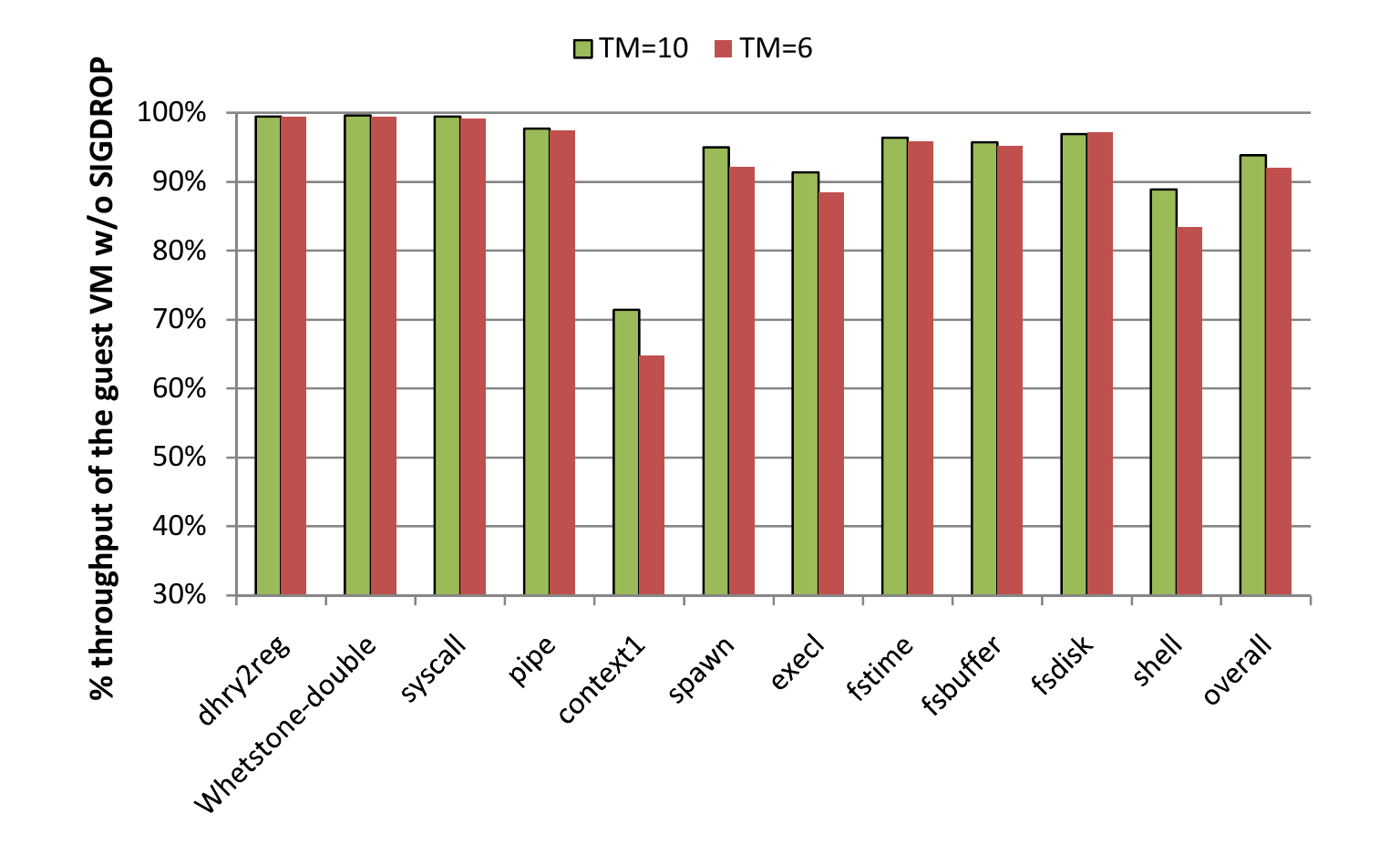}}
	\caption{The UnixBench benchmark results: system throughput degradation of a guest VM when SIGDROP is enabled with $T_M=10$ and $T_M=6$.}
	\label{fig:unixbench}
\end{figure}

{\bf UnixBench Benchmark} We choose UnixBench 5.1.3 \cite{unixbench} to test more aspects of the guest system's performance with and without SIGDROP. We perform 11 individual tests from the suite including process creation, pipe-based context switching and process communication, file copying, system call invocation, starting and reaping shell script, etc. The overall presents the average. Figure \ref{fig:unixbench} shows the results of the experiment averaged over 20 runs. When $T_M=10$ and $T_M=6$, the average system throughput degradation is 6.19\% and 7.94\%, respectively. 

\subsection{Storage Overhead}
Because SIGDROP detects ROP attacks based on general signatures that are independent of the monitored program, only the thresholds $T_M$ and $T_I$ need to be stored in the memory. One byte is enough for storing each threshold, which is negligible. Additionally, for each monitored hardware event, one bytes is used to store the occurrences (i.e., up to 256, which is large enough for the occurrences of any event within a monitor interval). With three events monitored simultaneously, each monitored process only requires 3 bytes of storage, which is also negligible. 

\section{Related Work}
\label{sec:prior}
KBouncer \cite{kbouncer} and ROPecker \cite{ropecker} use the last branch record (LBR) hardware registers to trace the target addresses
of indirect branches and compare them against the golden control flow path of the software. Since LBR registers are only available on Intel
platforms, KBouncer and ROPecker are not portable to AMD and ARM platforms. On the other hand, SIGDROP can be adapted to commodity platforms
with readily available HPCs.

ROPdefender uses dynamic binary instrumentation to verify the software CFI at the instruction trace granularity \cite{ropdefender}. For each $\mathtt{call}$ instruction to execute, ROPdefender pushes the address of its $\mathtt{return}$ instruction on a shadow stack. For each $\mathtt{return}$ instruction, ROPdefender pops the address on top of the shadow stack and compares it to the address on top of the program stack. ROPdefender incurs an average performance overhead of 2.17x for integer SPEC CPU 2006 benchmarks. SIGDROP has an overage overhead of 4.67\% for the same benchmarks.

G-Free is a compiler-based approach to thwart ROP attacks by eliminating gadgets in x86 binaries without altering legitimate software behavior \cite{g-free}. G-Free first protects the target address of aligned free-branch instructions\footnote{Free-branch instructions are unconditional indirect branch instructions.} by encrypting the target address of each $\mathtt{return}$ instruction, and by adding a random cookie on the program stack before the target address of each $\mathtt{jump}$ or $\mathtt{call}$ instruction. Second, G-Free removes unaligned free-branch instructions by replacing them with equivalent instructions. G-Free requires access to the software source code which may not be available. Moreover, the changes to the software binary may add new gadgets, defeating the purpose of the security countermeasure. SIGDROP is portable and induces no new gadgets because it doesn't modify the source code.

Davi et. al. propose two new processor instructions that enforce a golden CFI model \cite{hardware-cfi}. For each direct and indirect $\mathtt{call}$ instruction, a $\mathtt{cfibr\ [label]}$ instruction is added, where $\mathtt{[label]}$ is a hard-coded, unique immediate value associated to the  $\mathtt{call}$. $\mathtt{cfibr}$ pushes the label in a protected memory segment. If no $\mathtt{cfibr}$ is found for the  $\mathtt{call}$ instruction, the processor assumes a CFI violation. Each  $\mathtt{return}$ instruction has an associated $\mathtt{cfiret\ [label]}$ instruction that verifies if the $\mathtt{[label]}$  is in the protected memory segment. If no matching $\mathtt{[label]}$ is found, a CFI violation is detected. The proposed approach requires changes to different stages of the processor pipeline to incorporate the new instructions and is thus not portable to platforms currently available.

Reusing HPCs for security purposes has been applied to many defense mechanisms. Demme \textit{et al.} have demonstrated a techniqeu to use HPCs to detect Android malware and Linux rootkits \cite{jdemme-isca13}. Wang et. al. propose HPC-based runtime kernel rootkit detection and identification in a virtualization environment \cite{xwang2016tcad}. ConFirm is an HPC-based malicious firmware detection to secure embedded platforms with limited computing resources \cite{wang2015confirm}. Ozsoy \textit{et al.} have developed an HPC-based always-on hardware malware detection engine \cite{mozsoy-hpca15}. BRAIN leverages HPCs to measure the occurrences of low-level hardware events to detect Distributed Denial of Servic (DDoS) attacks \cite{jyothi2016brain}.

\section{Conclusion}
\label{sec:conclusion}
SIGDROP is a low-cost ROP detection approach which is based on low-level properties inherent to ROP attacks. Specifically, we observe special patterns in terms of certain hardware events when a ROP attack occurs during program execution. Such hardware event-based patterns form a signature to flag ROP attacks at runtime. SIGDROP leverages Hardware Performance Counters, which are already present in most commodity processors, to efficiently capture and extract the signature. We implement a prototype of SIGDROP on Linux. Our evaluation demonstrates that SIGDROP can effectively detect ROP attacks with acceptable performance overhead and negligible storage overhead.

\scriptsize

\bibliographystyle{IEEEtran}
\bibliography{sigdrop}

\end{document}